\documentclass[fleqn,10pt]{wlscirep}
\usepackage[utf8]{inputenc}
\usepackage[T1]{fontenc}
\usepackage{lineno}

\title{Super-resolution imaging for the detection of low-energy ion tracks in fine-grained nuclear emulsions}

\author[1,2,*]{Andrey ALEXANDROV}
\author[1,2]{Takashi ASADA}
\author[1]{Fabio BORBONE}
\author[2]{Valeri TIOUKOV}
\author[1,2]{Giovanni {DE LELLIS}}

\affil[1]{Università degli Studi di Napoli Federico II, I-80126 Napoli, Italy}
\affil[2]{I.N.F.N. sezione di Napoli, I-80126 Napoli, Italy}

\affil[*]{Corresponding author: andrey.alexandrov@na.infn.it}

\begin{abstract}
We propose a new  wide-field imaging method that exploits the Localized Surface Plasmon Resonance phenomenon to produce super-resolution images with an optical microscope equipped with a custom design polarization analyzer module. In this paper we describe the method and apply it to the analysis of low-energy carbon ion tracks implanted in a nuclear emulsion film. The result is then compared with the measurements of the same tracks carried out at an electronic microscope. The images set side by side show their close similarity. The resolution achieved with the current microscope setup is estimated to be about 50~nm. 
\end{abstract}
\begin{document}

\flushbottom
\maketitle
\thispagestyle{fancy}
\lhead{\textsc{Prepared for submission to Scientific Reports}}
\pagestyle{empty}

\section*{Introduction}

Nanomaterials are at the cutting edge of rapidly evolving nanotechnology. Because of their unique size-dependent qualities, these materials are employed in many applications.
Nanotechnology has the potential to provide information on the structure, function, and organization of a nanoscale system. 
Size-related features of nanoparticles (NPs) open up an endless number of possibilities for unexpected discoveries, while their often  unprecedented behavior pave the way for creative technological applications, but it also creates serious challenges for scientists. 
They must create highly controlled synthesis methods, more sensitive characterisation techniques, and, eventually, new models and theories to explain the experimental results.

The need for high-resolution imaging of objects at the nanometer scale with spatial resolution below the diffraction limit has given rise to a number of super-resolution methods. 
Many of them, including stochastic optical reconstruction microscopy (STORM)\cite{STORM}, spontaneous emission and photoactivated localization microscopy (PALM)\cite{PALM}, stimulated emission depletion (STED) microscopy\cite{STED} and super resolution by polarization demodulation (SPoD)\cite{SPoD}, make use of the optical characteristics of fluorescent emitters to selectively turn on or off nearby molecules or nanoparticles. 
Recent studies have shown that the localized surface plasmon resonance (LSPR) phenomenon may be used to image non-fluorescent metallic NPs in dielectric media using orientation-dependent localization microscopy (ODLM)\cite{ODLM} and super-resolution plasmonic imaging microscopy (SRPIM)\cite{SRPIM}.

Nuclear emulsion~\cite{Emul_GDL} is the oldest  nanomaterial used in nuclear and particle physics. It led to numerous discoveries since 1896 when Becquerel first observed the  radioactivity  until recently when  the discovery of tau neutrino appearance from a muon neutrino beam was reported by the OPERA experiment~\cite{OPERA_discovery,OPERA_final}. 
Nuclear emulsion is made up of nanoscopic silver bromide (AgBr) crystals suspended in a gelatin~\cite{Emul_GDL}. The ionization loss of a passing-through charged particle activates the crystals, which operate as sensors. The crystals' activated state is maintained until the emulsion film is chemically developed. As a result, a particle track is recorded, first as a sequence of activated crystals, then as a sequence of silver nanoparticles known as grains~\cite{Emu_track}. Depending on the emulsion type, these grains take the shape of randomly oriented filaments with typical diameters of several tens of nanometers~\cite{Emul_GDL}.

The Nano-Imaging Tracker (NIT)~\cite{NIT_Asada}, a novel type of fine-grained emulsion, was developed primarily for its use as a detector in the NEWSdm (Nuclear Emulsion WIMP Search with Directional Measurement) experiment~\cite{NEWS_EPJC}. This experiment develops the next-generation strategy for directional detection~\cite{Directional_review,Battat_Readout} of the so-called Weakly Interacting Massive Particle (WIMP)~\cite{WIMP_1,WIMP_2,WIMP_3} with a new and complementary approach intended to provide unambiguous signature of the galactic origin of dark matter. Nuclear emulsion show a very important feature for the investigation of the galactic origin of dark matter: the preservation by nuclear recoils of the direction of impinging dark matter particles~\cite{Directionality}.

Previously, two types of analysis were developed for short sub-resolution tracks appearing in NIT emulsions of the NEWSdm experiment: the elliptical fit (EF) of track images~\cite{ELLFIT} and the barycenter shift (BS) analysis of track images in response to the variation of the polarization also by the SRPIM method~\cite{SRPIM,SRMIC_2020}. 
The elliptical fit approach is rather simple and straightforward but its accuracy strongly depends on the perfectness of the optical conditions since even weak aberrations can spoil the angular resolution. Although the ellipticity parameter is correlated with the track length, a reliable measurement, however, is  only possible for tracks with lengths longer than 80\% of the microscope resolution. 
The barycenter shift analysis proved to be a rather good approximation for single- and two-grain tracks, which is expected to be the majority of events in the dark matter particle interactions. 
However, for events with more than two grains involved, the barycenter shift model rapidly becomes less accurate. 
Moreover, even in the two-grain case, the barycenter shift distance has only a weak correlation with the real distance between grains and can be considered as a lower limit of the track length rather than an estimate of the track length itself.   
Therefore, only the direction of nanometric tracks could be reliably measured by both methods mentioned above and none of them is able to measure the length of the track.
Fit models used by these methods (few grains along a straight line for the EF and two non-spherical grains with orthogonal orientations for the BS) provide reliable information about the inner structure of the track event only if the event configuration complies with the model. 
However, this is not always the case, thus, limiting strongly the efficiency.
Thus, both the EF and the BS types of analyses can reliably measure only the direction of sub-micron tracks, providing only limits on their lengths.

Precise measurements of the length of a track and the grain distribution inside it are important because they enable reconstruction of the energy of the charged particle as well as its mass, if necessary. This is a challenging task for a conventional optical microscope due to its diffraction-limited resolution. Although this kind of measurements can be performed with a higher-resolution device, such as an X-ray microscope or a Scanning Electron Microscope (SEM), the data acquisition in both cases would be much slower than with an optical microscope. Moreover, emulsion samples in the SEM case would require depositing a conductive coating and only few hundreds of nanometers near the sample surface would be accessible.

In this work we propose a new fast super-resolution imaging technique that can be used with optical microscopes.
It exploits the polarization dependency of the LSPR phenomenon, but unlike the SRPIM, is capable of reconstructing images of nanoparticles. 
Another advantage of the proposed method is that it requires a limited number of images taken at different polarization, eight images in our case, that technically can be taken simultaneously thus making possible super-resolution imaging in real-time. 

\section*{Results}

\subsection*{Plasmon resonance and silver grains}
Localized surface plasmon resonance is an optical phenomenon occurring when a light wave, trapped within conductive nanoparticles with dimensions smaller than the wavelength of light, interacts with the surface electrons in the conduction band. 
This interaction produces coherent localized plasmon oscillations with a resonant frequency that strongly depends on the composition, size, geometry, dielectric environment and separation distance of NPs. 
The interaction of NPs with light allows some photons to be absorbed and some others to be scattered. The scattering cross-section is given by the formula~\cite{LSPR_Myrosh}:

\begin{equation} \label{eq:scatCS}
\sigma^{sc} = \frac{8\pi^{3}}{3\lambda^{4}}|\alpha|^{2},
\end{equation}

where $\lambda$ is the wavelength of the incident light and $\alpha$ is the polarizability. The polarizability represents a distortion of the electron cloud in response to an external electric field. NPs smaller than about 150 nm respond basically as induced dipoles~\cite{LSPR_Myrosh}, therefore, neglecting higher orbital moments, the polarizability of a small ellipsoid along one of the principal axes $j=x,y,z$ takes the form:

\begin{equation} \label{eq:AlphaJ}
\alpha_{j} = V\epsilon_{m}\frac{\epsilon-\epsilon_{m}}{\epsilon_{m}-L_{j}(\epsilon-\epsilon_{m})},
\end{equation}

where $V$ is the particle volume, $\epsilon=\epsilon_1+i \epsilon_2$ and $\epsilon_m$ are (complex) permittivities of the NP and of the surrounding medium, respectively, and $L_j$ are depolarization factors satisfying the condition $L_x+L_y+L_z=1$.
Thus, putting together the equations \ref{eq:scatCS} and \ref{eq:AlphaJ} one gets for the scattering cross section along the $j$-th axis:

\begin{equation} \label{eq:scatCS-ell}
\sigma_{j}^{sc}=\frac{8\pi^{3}V^{2}\epsilon_{m}^{2}}{3\lambda^{4}}A_{j}^{2}\frac{(\epsilon_{1}-\epsilon_{m})^2+\epsilon_{2}^{2}}{(\epsilon_{1}+\chi_{j}\epsilon_{m})^2+\epsilon_{2}^{2}},
\end{equation}

where the factors $\chi_{j}=(1-L_{j})/L_{j}$ and $A_{j}=1/L_{j}$ determine the plasmon resonance peak wavelength and the polarization pattern of the scattered light, respectively.

It can be shown~\cite{LSPR_Petryaeva} that for elongated NPs, with the longest dimension  parallel to the $z$ axis, the depolarization factors depend on the aspect ratio (also called the ellipticity) $R=length/width$ as

\begin{equation} \label{eq:depolFact}
L_{z} = \frac{1-e^2}{e^2}[\frac{1}{2e}\ln{(\frac{1+e}{1-e}})-1];\\L_{x}=L_{y}=\frac{1-L_{z}}{2};
\end{equation}

where $e^{2}=1-(\frac{width}{length})^2=1-R^{-2}$.

For elongated NPs the depolarization factor $L_{z}$ is always smaller than $1/3$ and, therefore, it is always $A_{z}>A_{x,y}$, meaning that the scattered light gets polarized in the direction of the longest dimension of the NP. For this reason, the intensity of the light scattered from elongated NPs and observed at an optical microscope equipped with a rotatable polarizer plate changes between two extreme values,  depending on the aspect ratio and on the orientation of the NP with respect to the optical axis. This intensity is a periodic function of the polarization angle $\theta$ that can be described with the cosine law:

\begin{equation} \label{eq:I-theta}
I_{\theta}=a \cos{(2[\theta-\phi])} + b,
\end{equation}

where $\phi$ is the polarization angle where the intensity is maximal, $a$ and $b$ are the amplitude and the mean value of the intensity, respectively. The factor $2$ in Eq. \ref{eq:I-theta} accounts for the equality of polarization angles rotated by $\pi$. 

This behaviour paves the way for resolving closely-located NPs, even at distances shorter than the diffraction limit. Indeed, except the very unlikely configuration where both the shape and the orientation of NPs are identical, variations of the intensity of the scattered light from each NP are different, thus producing variations of the image of unresolved NPs. By analysing the variations of the image shape and wavelength spectrum, one can deduce the number of NPs, their dimensions, shape and orientation. In this work we focus on exploiting the variations of the image shape as a function of the polarization angle. 

\subsection*{Super-resolution imaging method}
\begin{figure}
\centering
{\includegraphics[width=0.8\columnwidth]{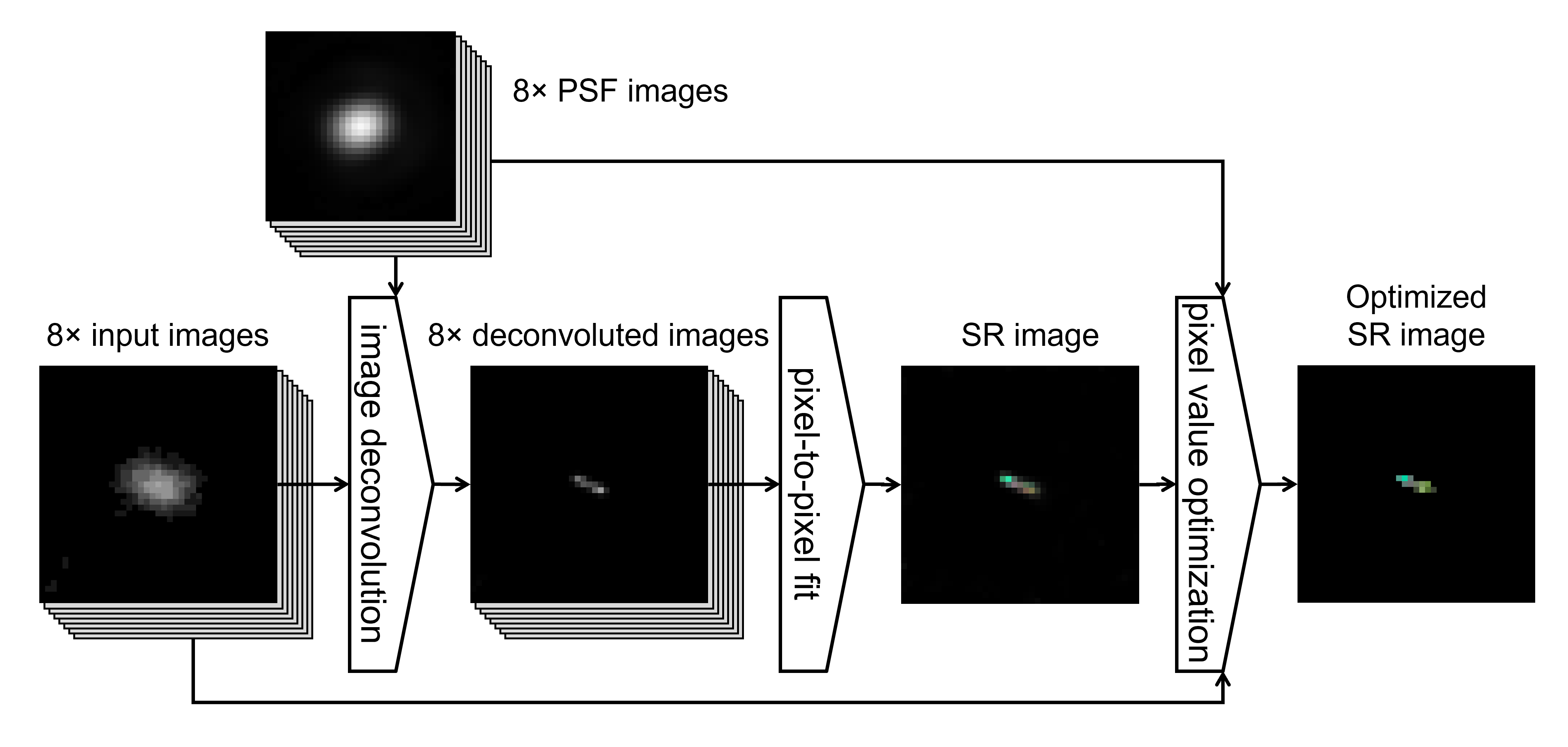}}
\caption{ The scheme of the proposed super-resolution imaging method.
}
\label{fig:Figure1}
\end{figure}

\begin{figure}
\centering
{\includegraphics[width=0.9\columnwidth]{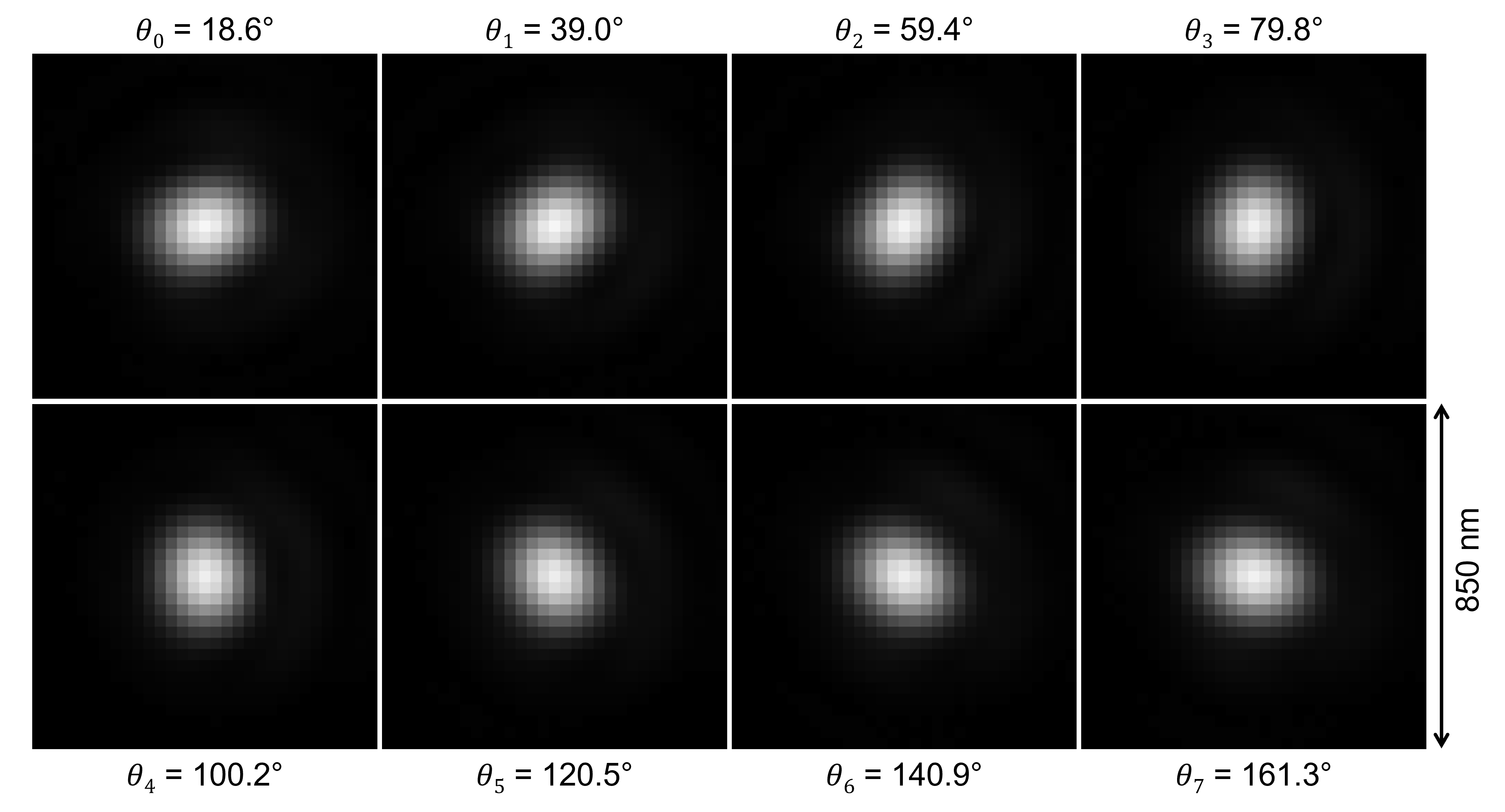}}
\caption{PSF images for different polarization angles. Single image dimensions: 31 $\times$ 31 pixels approximately corresponding to 850 nm $\times$ 850 nm. The values of the polarization angles are reported for each image.
}
\label{fig:Figure2}
\end{figure}

\begin{figure}
\centering
{\includegraphics[width=0.9\columnwidth]{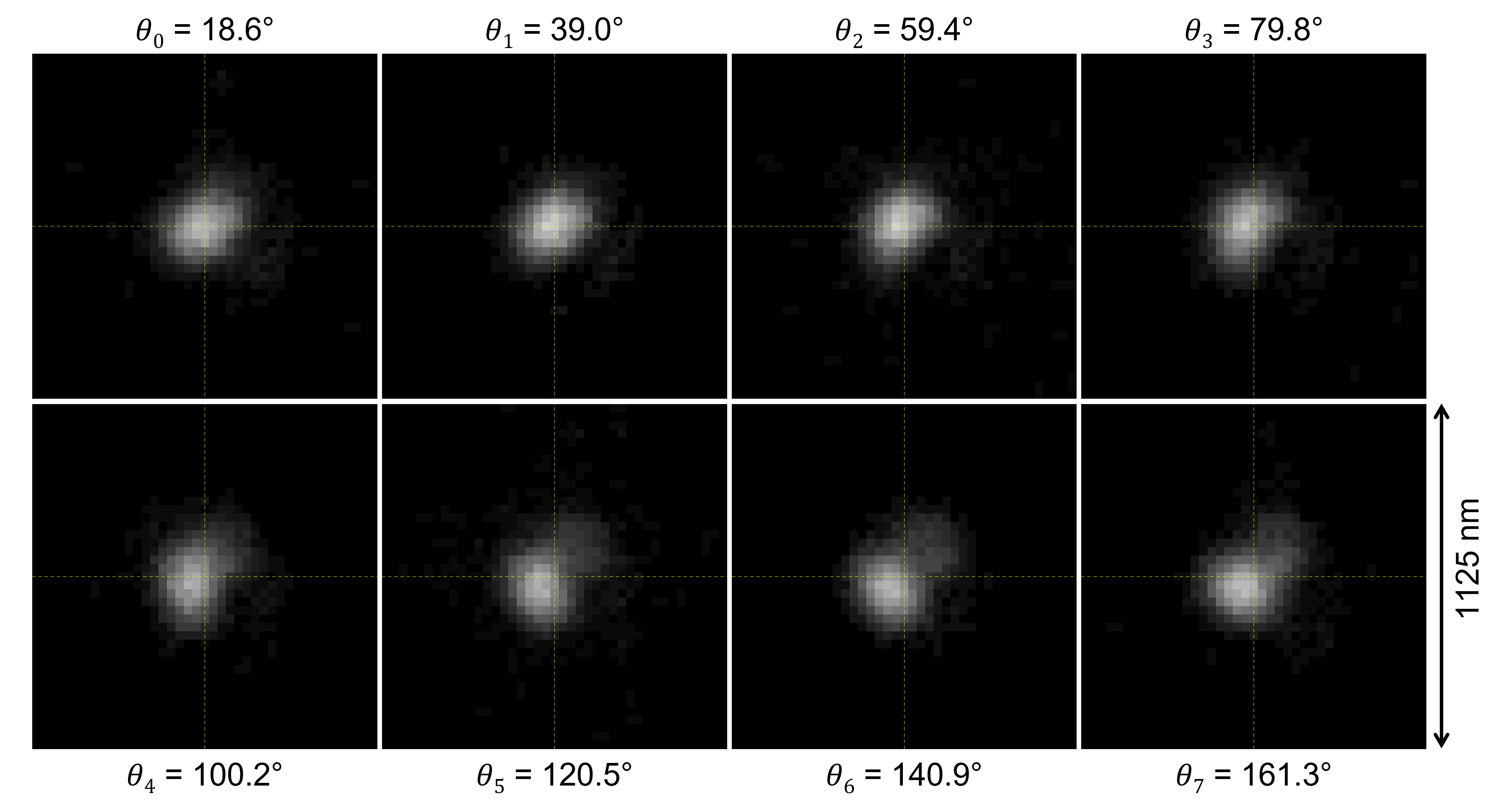}}
\caption{Microscopic images of a 100 keV Carbon ion track taken at different polarization angles. Single image dimensions: 41 $\times$ 41 pixels approximately corresponding to 1125 nm $\times$ 1125 nm. The values of the polarization angles are reported for each image.
}
\label{fig:Figure3}
\end{figure}

The scheme of the imaging method is shown in Figure~\ref{fig:Figure1}. It takes images of the same event at different polarization angles. In the present study the number of images was chosen to be 8, but, in general, it can be arbitrary. The knowledge of the microscopic point spread function (PSF) at each polarisation angle is needed: it can be measured directly at the microscope or generated by an analytical solution.
In general, as shown in Figure~\ref{fig:Figure2}, PSFs corresponding to different polarization angles are slightly different. Indeed, after the rotation of the polarization angle, the imaging system changes its optical properties and each polarization angle has to be considered as a separate optical system with own PSF.
Hence, images taken at a given polarization angle are processed with the PSF associated to that polarization angle. An example of microscopic images of an event taken at different polarization angles is shown in figure~\ref{fig:Figure3}. The polarization-dependent plasmon resonance effect is clearly visible by observing changes in the image shape and its barycenter position.

The imaging method comprises three steps described in the following of this section: image deconvolution, pixel-to-pixel fit and pixel value optimization.

\paragraph{Image deconvolution.}
The knowledge of the PSF allows applying numerical deconvolution methods in order to correct the distortion of images taken at  optical microscopes. We have developed a deconvolution method that implements the Maximum Likelihood Expectation Maximization (MLEM) algorithm~\cite{MedImg_book}.
The advantage of this approach is that the MLEM method does not require the estimation of derivatives of the likelihood function. The solution is approximated through an iterative procedure:

\begin{equation} \label{eq:MLEM-iter}
x_{i}^{(n+1)} = x_{i}^{(n)}\cdot\frac{1}{\sum_{j} A_{ij}}\cdot\sum_{j}A_{ij}\frac{y_{j}}{\sum_{k}A_{kj}x_{k}^{(n)}},
\end{equation}
where $y_{j}$ are the pixel values of the observed image, $x_{i}^{(n)}$ are pixel values of the deconvoluted image in the $n$-th iteration and $A_{ij}$ is the convolution matrix that describes the optical system. The matrix $A_{ij}$ is constructed using the PSF to hold the equivalence $y_{j}^{(n)}=PSF\bigotimes x_{k}^{(n)}\equiv\sum_{k}A_{kj}x_{k}^{(n)}$, where the image $y_{j}^{(n)}$ is the result of the convolution of the image $x_{k}^{(n)}$ with the PSF.

\begin{figure}
\centering
{\includegraphics[width=0.9\columnwidth]{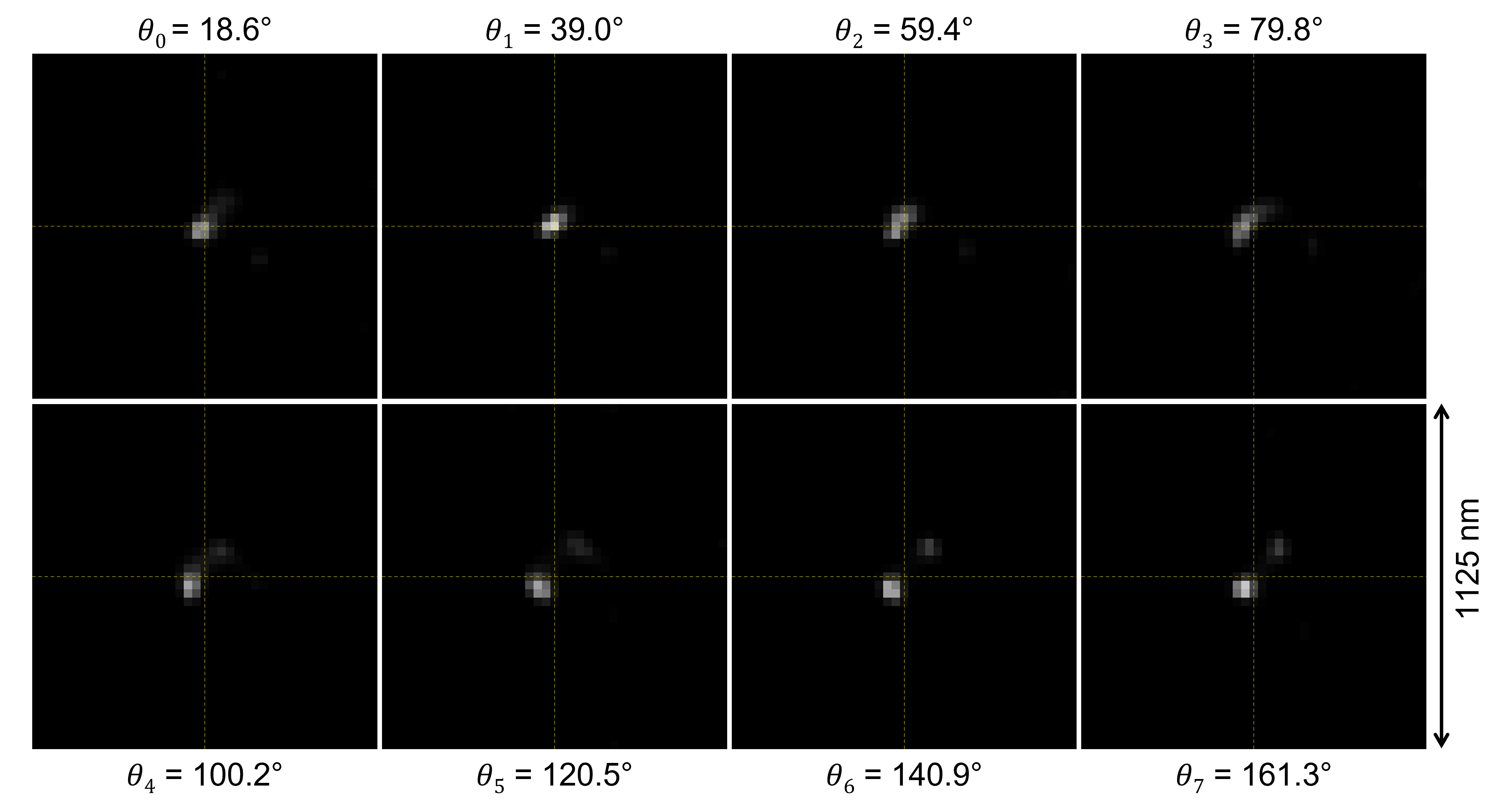}}
\caption{Deconvoluted images of the same event reported in Figure~\ref{fig:Figure3}. Single image dimensions: 41 $\times$ 41 pixels approximately corresponding to 1125 nm $\times$ 1125 nm. The values of the polarization angles are reported for each image.
}
\label{fig:Figure4}
\end{figure}

The deconvolution of the images reported in figure~\ref{fig:Figure3} is shown in figure~\ref{fig:Figure4}. It can be noted that the diffraction-limited spots are reduced to much smaller "cores" of just 1 or 2 pixels in width. These cores slightly change their shape and position as a function of the polarization. Moreover, a new core appears in the images corresponding to the last four polarization angles. We interpret these cores as images of actual grains in the emulsion. 

\begin{figure}
\centering
{\includegraphics[width=0.9\columnwidth]{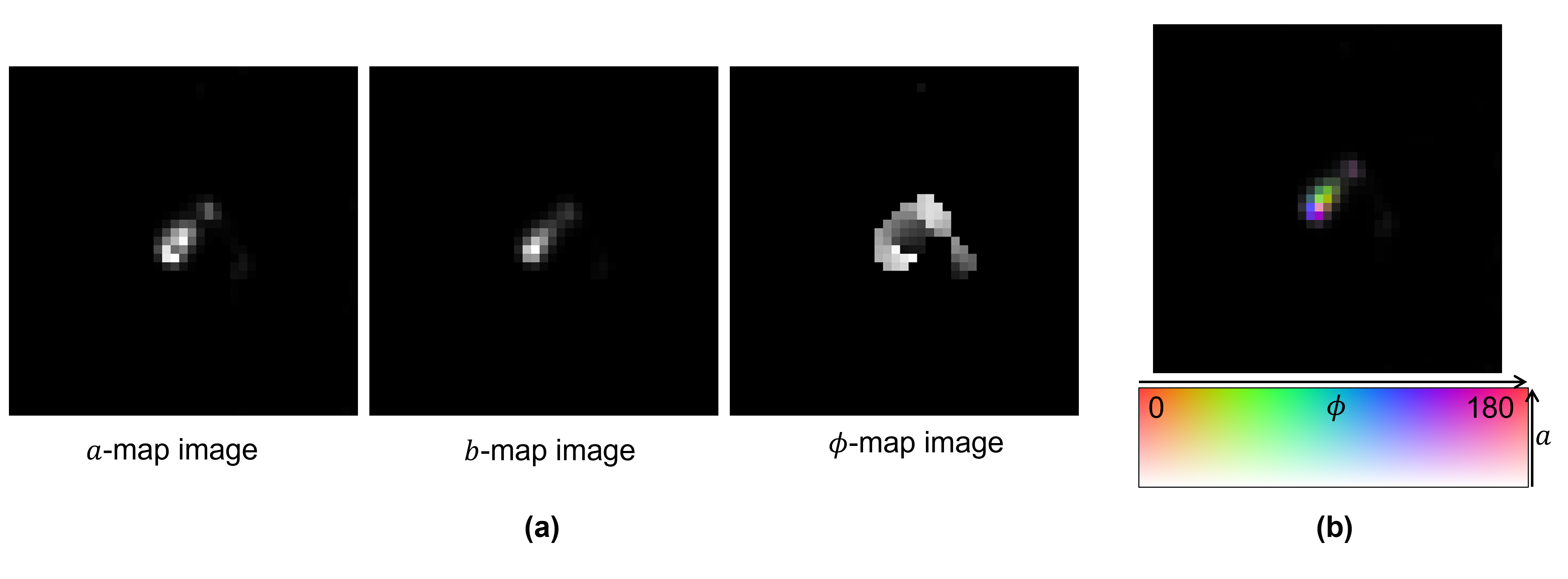}}
\caption{Pixel-to-pixel fit result for the same event reported in figure~\ref{fig:Figure3} and ~\ref{fig:Figure4}. (a) $a$-, $b$- and $\phi$-map representation. (b) HSV color space representation. Single image dimensions: 41 $\times$ 41 pixels approximately corresponding to 1125 nm $\times$ 1125 nm.
}
\label{fig:Figure5}
\end{figure}

\paragraph{Pixel-to-pixel fit.}
The modification of the core position as a function of the polarization angle visible in figure~\ref{fig:Figure4} indicates the presence of yet unresolved close grains. The scattering intensity modulation described by equation (\ref{eq:I-theta}) makes the scattering center to move from one grain to another. Without loss of generality, we assume that every pixel in deconvoluted images is modulated according to the same function in Eq.~\ref{eq:I-theta}: 

\begin{equation} \label{eq:pix-mod}
I_{i,k}=a_{i} \cos{(2[\theta_{k}-\phi_{i}])} + b_{i},
\end{equation}

here $I_{i,k}$ is the intensity of the pixel $i$ in the image $k$ taken at the polarization angle $\theta_{k}$. Unknown parameters $a_{i}$, $b_{i}$ and $\phi_{i}$ have the same meaning as parameters $a$, $b$ and $\phi$ in Eq.~\ref{eq:I-theta} and can be found by fitting values $I_{i,k}$ with the function in Eq.~\ref{eq:pix-mod} for all $k$. The fit result for images from figure~\ref{fig:Figure4} is shown in figure~\ref{fig:Figure5} by means of two different representations. 

The three plots in figure~\ref{fig:Figure5}a) show the result of the fit in the form of three separate images: $a$-map, $b$-map and $\phi$-map, each consisting of the corresponding estimated $a_{i}$, $b_{i}$ and $\phi_{i}$  parameters, respectively. In this representation, the range of the parameters is transported into the 8-bit greyscale range. This is particularly true for the phase values ranging between $0$ and $\pi$. It is worth noting that the $\phi$-map pixel values undergo a sharp discontinuity at 0 and $\pi$ phases, having the most distant greyscale values (black and white) despite being physically the same polarization angle.

In order to overcome this limitation, we have adopted an alternative representation shown in figure~\ref{fig:Figure5}b, where we used the Hue-Saturation-Value (HSV) encoding~\cite{HSV}, by setting for the $i$-th pixel: $hue_{i}=2\phi_{i}$, $saturation_{i}=a_{i}$ and $value_{i}=(a_{i}+b_{i})$. 
$Hue$ is the color space; $Saturation$ measures the departure of $hue$ from achromatic, i.e., from white or gray. $Value$ measures the departure of $hue$ from black.
The color palette in the bottom of  figure \ref{fig:Figure5}b shows the hue-saturation plane for $Value = 1$. The horizontal axis represents hue values corresponding to different colors, while the vertical axis is the saturation showing how a certain $hue$ changes from  white in the bottom ($saturation=0$) to a strong color in the top.
The HSV representation is typically more convenient since it combines all the parameters in an intuitive way: the color indicates the pixel phase, the color saturation is equal to the modulation amplitude and the brightness corresponds to the maximal pixel brightness during the modulation. It must also be noted that in this representation phases 0 and $\pi$ have exactly the same color (red in the color palette of the figure \ref{fig:Figure5}b) and, hence, there is no problem of discontinuity. Thus, every pixel in a HSV-encoded image has three dimensions and, therefore,  can also be represented by three images called channels: hue channel, saturation channel and value channel. Hue and saturation channels correspond to $\phi$-map and $a$-map images, respectively, while the value channel is the sum of $a$-map and $b$-map images.

\begin{figure}
\centering
{\includegraphics[width=0.7\columnwidth]{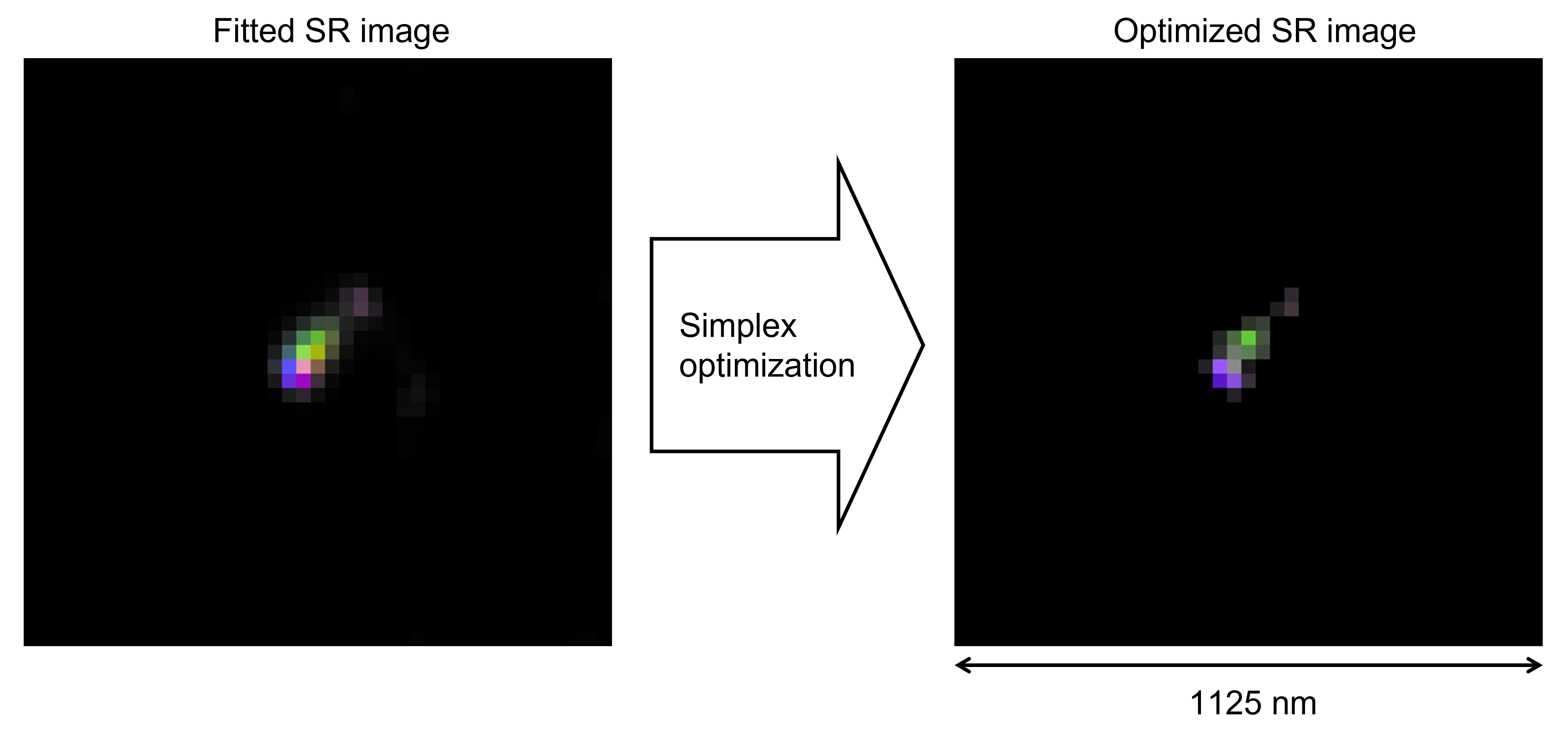}}
\caption{ (left) SR-image after pixel-to-pixel fit. (right) Optimized SR-image. The same event reported in  figure~\ref{fig:Figure3} and~\ref{fig:Figure4} is shown here. Single image dimensions: 41 × 41 pixels approximately corresponding to 1125 nm × 1125 nm.
}
\label{fig:Figure6}
\end{figure}

\paragraph{Pixel value optimization.}
Since the pixel-to-pixel fit function, given by equation~\ref{eq:pix-mod}, does not consider values of neighboring pixels, the noise in pixel brightness values affects the accuracy of the fit, especially in the phase-space. In order to fix that, an optimization step was introduced. In this step  the $a_{i}$, $b_{i}$ and $\phi_{i}$ pixel values of the SR-image are used to calculate pixel brightness $I_{i,k}$ at a specific polarization angle $\theta_{k}$. Then, the image $I_{i,k}$ is convoluted with the PSF of that polarization angle and the obtained image $x_{i,k}$ is compared with the image $y_{i,k}$ observed at the microscope at that polarization angle. The residual $r_{k}$ is calculated as

\begin{equation} \label{eq:residual}
r_{k}=\frac{1}{X_{k} Y_{k}}\sum_{i}\frac{(X_{k} y_{i,k} - Y_{k} x_{i,k})^{2}}{x_{i,k} + y_{i,k}},
\end{equation}

where $x_{i,k}=\sum_{j}A_{ij,k} I_{j,k}$, $X_{k}=\sum_{i}x_{i,k}$ and $Y_{k}=\sum_{i}y_{i,k}$.
The same procedure is repeated for all remaining polarization angles and a total residual is calculated as $r=\max_{k}(r_{k})$. 

Then, the residual value $r$ is minimized numerically following the Simplex method described, for example, in Ref.~\cite{Simplex_book} by adjusting values $a_{i}$, $b_{i}$ and $\phi_{i}$ of the SR-image with the pixel-to-pixel fit result taken as initial value.  Since the calculation of the convolution with a PSF involves neighboring pixels as well, the noise effect is reduced and sharper SR-images with more distinct color (same pixel phase) cores are produced, thus indicating distinct scattering centers. The result of the SR-image pixel optimization is shown in figure~\ref{fig:Figure6}, where three aligned scattering centers are clearly visible, denoted in violet, green and violet again colors. The latter center is noticeably weaker than the other two and, therefore, it seems grey due to small saturation and brightness values.

\subsection*{Comparison with measurements at SEM}

\begin{figure}
\centering
{\includegraphics[width=0.9\columnwidth]{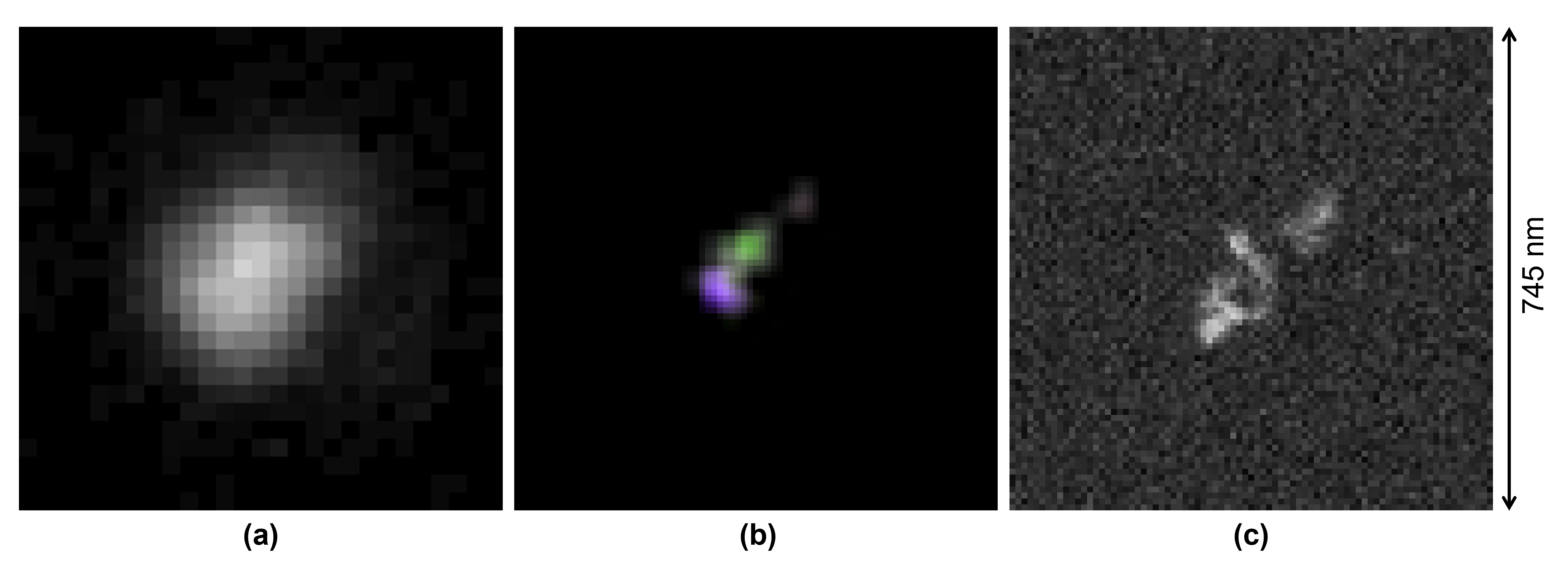}}
\caption{(a) 100 keV carbon ion track image at a conventional optical microscope, (b) SR-image after 3$\times$ bicubic interpolation and (c) SEM image of the same track. Image dimensions: (a) 27$\times$27 pixels, (b) and (c) 81$\times$81 pixels. All images approximately correspond to an area of 745 nm $\times$ 745 nm. The pictures show the same event reported in figure~\ref{fig:Figure3} and~\ref{fig:Figure4}.
}
\label{fig:Figure7}
\end{figure}

The described SR-imaging method was applied to 104 carbon ion tracks, implanted with a kinetic energy of 100 keV in an NIT emulsion film. 
The same emulsion film was also analysed with a SEM. 
Both microscopes have sufficiently large fields of view, also similar in dimensions (about 60$\times$45 $\mu$m), each containing tens of tracks, which allows application of pattern matching for unique track identification.
The events matching procedure is described in Supplementary materials.
It must be noticed that 100 keV Carbon ion tracks implanted in an emulsion film produce tracks with an average length of 300 nm and about 16\% of the tracks are shorter than 200 nm. 
A direct comparison between SR and SEM images was not possible due to the difference in pixel size: 9 nm/pixel for SEM images and 27.5 nm/pixel for SR-images, which results in a different number of pixels in images of the same surface. 
Therefore, we applied a 3$\times$ bicubic interpolation procedure~\cite{Bicubic} to  SR images, and made the number of pixels in the two types of images comparable. 
Moreover, SEM images contain only the information about intensities. 
For this reason, SR-images are ''depolarized'' by extracting the value channel prior to the comparison.

\paragraph{Similarity test.}
Plots in figure~\ref{fig:Figure7} show the same 100 keV Carbon ion track observed with a conventional optical microscope (a), the reconstructed SR-image after the interpolation (b) and the image obtained at the SEM (c), respectively. No detail is perceivable with the conventional optical microscope, while a three-grain structure is clearly visible in both SR and SEM images. Moreover, different colors in the SR-image indicate two adjacent grains, thus, providing one more lever arm to isolate grains even if they are too close to be resolved or if their images overlap.

\begin{figure}
\centering
{\includegraphics[width=0.5\columnwidth]{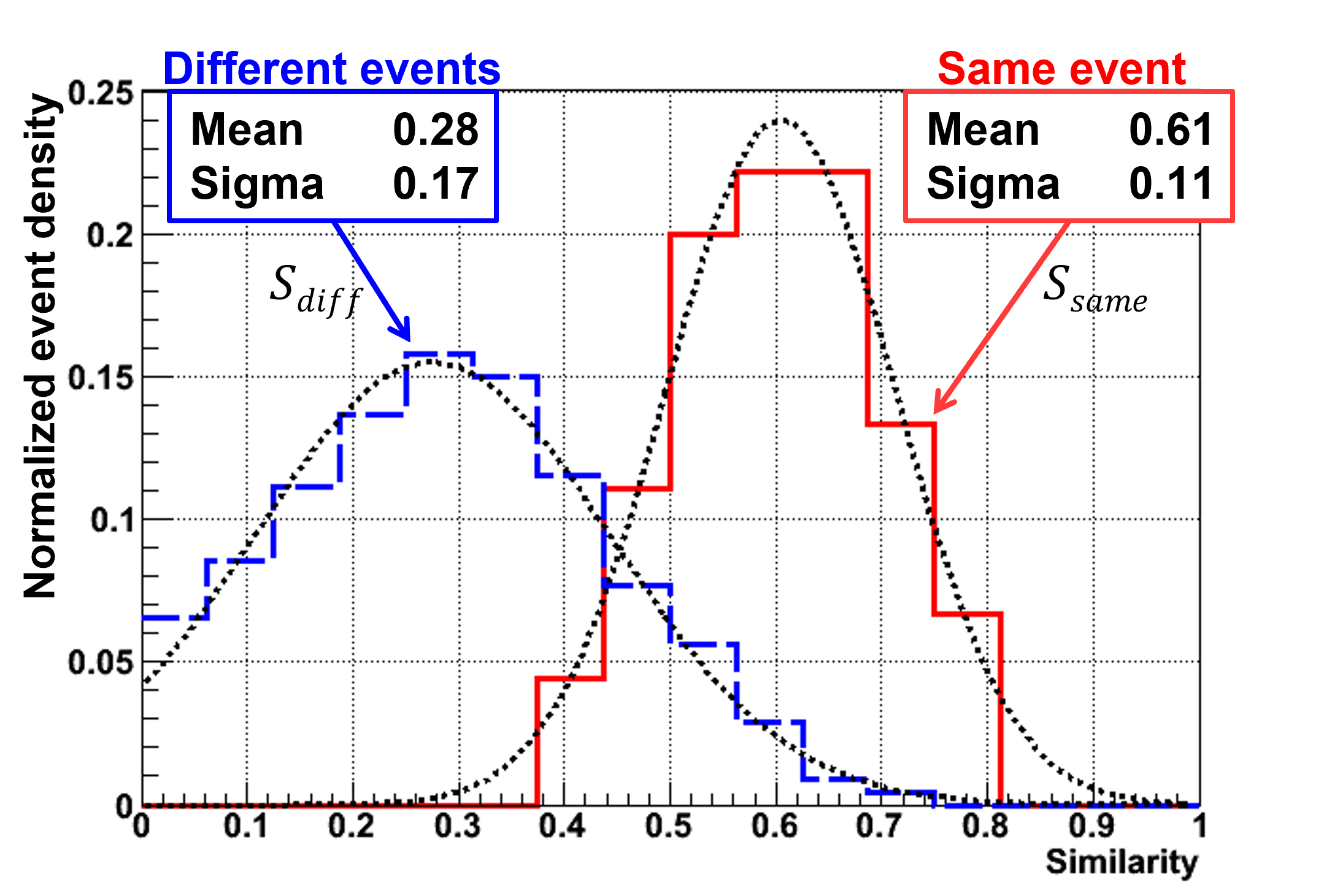}}
\caption{Similarity distribution for SR-SEM image pairs belonging to same tracks (red solid line) and to different tracks (blue dashed line). Both distributions are normalized to the unity of their integral.
}
\label{fig:Figure8}
\end{figure}

To quantify the similarity level between SR and SEM images we use the Pearson correlation function:
\begin{equation} \label{eq:SIM}
S=\frac{\sum_{i}(SR_{i}-\overline{SR})(SEM_{i}-\overline{SEM})}{\sqrt{\sum_{i}(SR_{i}-\overline{SR})^{2}}\sqrt{\sum_{i}(SEM_{i}-\overline{SEM})^{2}}},
\end{equation}
where $\overline{SEM}$ and $\overline{SR}$ are the mean pixel brightness of the SEM image and of the value channel of the SR-image, respectively. $S=0$ corresponds to no correlation between images while $S=1$ indicates that the two images are identical.
For all SR-SEM image pairs belonging to the same tracks we have estimated their similarity values $S_{same}$. After that, similarity values $S_{diff}$ were calculated for image pairs where SR and SEM images are taken from different tracks. The obtained distributions, after normalization, are shown in figure~\ref{fig:Figure8}. It can be seen that, on average, $S_{same}$ is twice higher than $S_{diff}$ confirming the correctness of the SR-image reconstruction. The relatively high $S_{diff}$ value can be explained by the similar structure of all events: bright spots near the image center. It must also be noted that we do not expect perfect identity between SR and SEM images of the same event since they are obtained from very different microscopes and special conditions had to be applied for the scanning at SEM. In particular, scanning at SEM required cleaning the sample from immersion oil, during which grains that are close to the surface can potentially be washed out. Moreover, scanning at SEM with a relatively energetic electron beam can potentially disrupt the interactions between the grain and the surrounding gelatin matrix, which can lead to their displacement from the place where they were during scanning at the optical microscope. The necessity of covering the sample with a conductive coating before scanning at SEM makes it impossible to rescan it at an optical microscope.

\paragraph{Estimation of the achieved resolution.}
The comparison of figures~\ref{fig:Figure7}b and~\ref{fig:Figure7}c shows that the shapes of individual grains are not perfectly reconstructed by the SR method indicating that the achieved resolution is still not as high as that of the SEM.  
The study with simulated images has shown that the convolution of a simulated image with realistic PSFs shown in figure~\ref{fig:Figure2} with the following application of the reported super-resolution method results in images identical to the original ones, which implies that the current resolution of the method is instrumentally limited.
The analysis shows that the resolution is limited by the camera sensor pixel size, which is also evident from figure~\ref{fig:Figure6}, where images before 3$\times$ interpolation are shown.



The achievable resolution of the described method can be estimated by using events simulation as described in the Supplementary material.
Simulation of the system's response to a point-like source provides an estimate of the resolution as $(45\pm 12)$~nm. 
Estimation of the distance at which two close point-like sources start being visible as separate ones provides an estimate of the resolution estimation of about 50 nm.
Both estimations are compatible and, to be conservative, we use the larger one.

\paragraph{Track length and direction measurement.}
Position resolution, track length and direction determination are the most important parameters for a tracking device. Moreover, they are particularly relevant  for the directional dark matter search conducted by the NEWSdm experiment: the track length is directly connected with the recoil energy while the nuclear recoil direction is highly correlated with the incoming direction of the dark matter particle. 
To define the track direction we first draw the best fit line passing through all pixels of the track image. The weight of a pixel in the fit is equal to its brightness for SEM images and to the intensity of the pixel $value$ parameter for SR-images. Then, we define the track direction as the angle of the best fit line with respect to the $X$ axis. After that, all pixels with the weight above a certain threshold are projected onto the best fit line an the track length is defined as the distance between the two extreme projections. The same procedure is applied for both SR and SEM images. Unlike SR-images where the background noise is negligible, a brightness cut is applied to suppress the background noise in SEM images.

\begin{figure}
\centering
{\includegraphics[width=0.8\columnwidth]{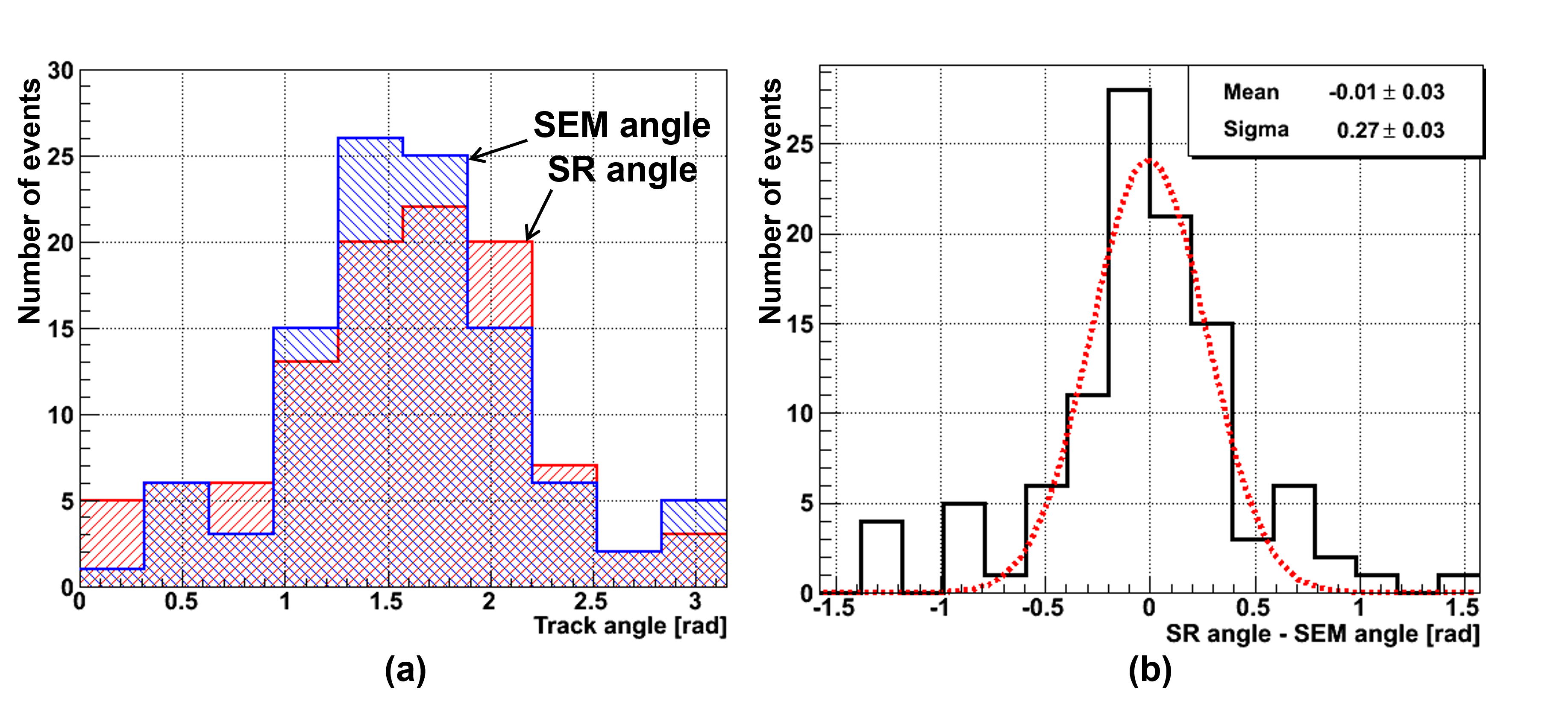}}
\caption{(a) Angular distribution of 100 keV carbon ion tracks reconstructed with the SEM (blue top-left-bottom-right diagonally hatched area) and with the SR (red top-right-bottom-left diagonally hatched area) microscope images. The peak indicates the known direction used for the implantation of these ions in the film. (b) Angular difference for tracks reconstructed by SR and SEM images.
}
\label{fig:Figure9}
\end{figure}

The angular distributions of the same 100 keV Carbon ion tracks reconstructed in SR and SEM images are shown in figure~\ref{fig:Figure9}a. The peak coincides with the known direction used for the implantation of these ions in the film. Some misalignment between peaks is due to the manual setting procedure of NIT film on the stages of both microscopes. The events at large angles with respect to the beam form the pedestal of the angular distribution. They are mainly due to single-grain events with a small $length/width$ ratio. These events are not produced by beam ions but are, rather, the result of the thermal activation of ArBr crystals. After the film development, the thermal excitation effect produces spuriously distributed silver grains, called fog.

Figure~\ref{fig:Figure9}b shows the angular differences of the same event track reconstructed by SR and SEM images. For most of the events, the angular accuracy is about $(270\pm30)$ mrad. A few events in tails are also due to short single-grain events. Since the instrumentally limited resolution is comparable with the grain size itself, the grain shape is not perfectly reconstructed, which decreases the accuracy of grain direction measurement. As it was mentioned before, there are also other reasons, like an intense electron beam, which can disrupt the interactions between the grain and the surrounding gelatin matrix, leading to displacement or rotation of the grain. Particularly high electron currents can, potentially, change the grain shape itself by heating it above the silver melting point. In this work a special care was taken to avoid grain overheating during measurements at SEM (see Methods section).

\begin{figure}
\centering
{\includegraphics[width=0.8\columnwidth]{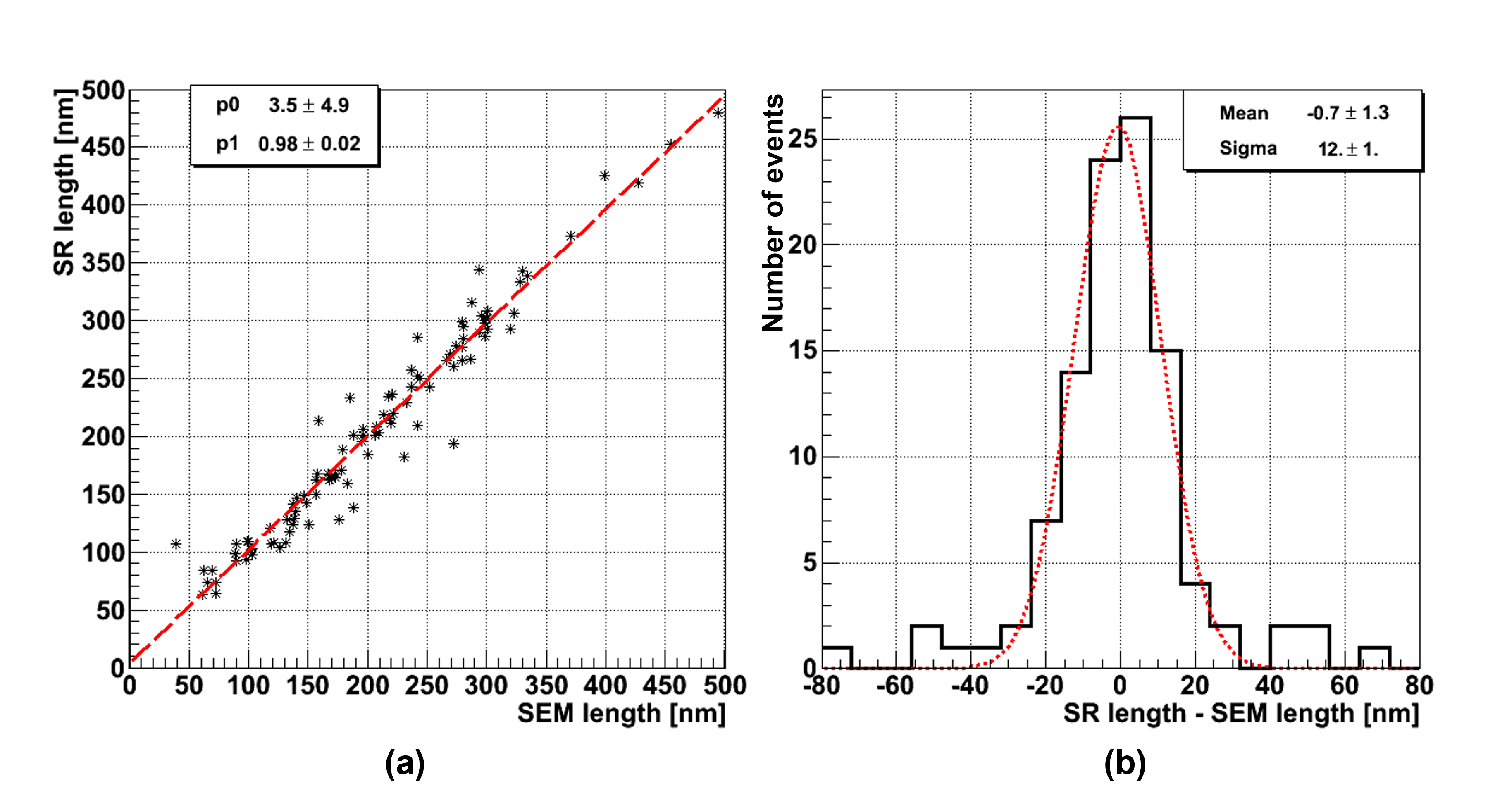}}
\caption{(a) Scatter plot of the lengths of 100 keV Carbon ion tracks measured with the SR (vertical axis) and the SEM  (horizontal axis) microscope. The red dashed line is the best fit. (b) Distribution of the difference between the lengths measured with the two methods. 
}
\label{fig:Figure10}
\end{figure}

As it is shown in figure~\ref{fig:Figure10}a, the correlation coefficient between track length measurements  by SR and SEM images is very close to 1 in the range between 60 nm and 500 nm. Figure~\ref{fig:Figure10}b shows the lengths difference. The distribution shows a Gaussian shape, to a large extent, with a sigma of about $(12\pm 1)$ nm. The tails in the distribution are mainly due to events with unmatched number of grains. As it was already mentioned, a loss of a close-to-surface grain can happen during the NIT film cleaning procedure in order to remove immersion oil residuals. Another possibility is the difference in the depth accessible to each microscope: the SEM is limited to about 200 nm from the surface while an optical microscope can access much deeper layers. This can also lead to situations when a grain is visible in the optical microscope but is missing at the SEM.

\section*{Discussion}

The  super-resolution method reported in this paper is based on the analysis of the polarization of the light scattered off metallic NPs immersed in a dielectric medium. The LSPR phenomenon introduces a noticeable anisotropy if the light scatters off non-spherical NPs. The polarization pattern is analysed by means of a specially designed optical microscope equipped with a polarization analyser. Super-resolution images reconstructed by the presented method show a high similarity level when compared with images of the same events observed at the SEM. 
This enables such optical microscopes to perform measurements at the nanoscale level, previously inaccessible due to the diffraction limit. It was also demonstrated that the length and the direction of ultra-short carbon ion tracks in emulsion can be measured with an accuracy comparable to that of electronic microscopes, but much faster: in this study it took about 1 minute to acquire one field of view at the SEM, while only 1 second was needed for a field of view of the same surface at the optical microscope, and faster implementations are envisaged. 

The proposed super-resolution method can be extended to all three dimensions by combining images taken at different depths or by introducing an additional vertical focal plane~\cite{Patent_mic}. 
A new 3D super-resolution optical microscope, orders of magnitude faster than the actual one, is now under construction at the University of Naples. 
Its unique design will enable the simultaneous capture of polarization images, allowing the use of novel scanning techniques\cite{LASSO_CM,LASSO_IM} and dramatically increasing readout speed.

The development of a fast wide-field super-resolution technique capable of reconstructing low-energy sub-micron ion tracks in nano-grained emulsions paves the way for the development of a solid-state detector that can identify WIMP events even
beyond the so-called neutrino floor where neutrinos would become an otherwise irreducible background source~\cite{NEWS_EPJC}.

Besides the directional dark matter search, the presented super-resolution imaging technique could be interesting for study of the fragmentation in proton-Nucleus (p-N)~\cite{FOOT_ref} collisions important for the hadrontherapy and for the radioprotection in space, which is an exceptionally challenging task because of the very short range of the produced fragments. Their range is limited to tens of microns and even a very thin solid target would stop them or badly spoil their energy measurement. Nowadays, existing experiments are forced to adopt the inverse kinematic approach to solve that problem, while with the proposed imaging technique direct measurements become possible.

In general, the proposed imaging technique can be employed for quick study of materials having anisotropic optical properties with respect to the polarization direction. 
It may, for example, be used to reconstruct the spatial orientation of NPs, which is crucial for understanding some important biological phenomena like the heterogeneous structure deformations of the cell membrane or the interaction force between proteins and cell membranes~\cite{APP1}. 
One possibility is to use it in microfluidics to visualize the orientation of non-spherical metallic NPs inside capillaries and near membrane walls, where the diffusive behavior of a NP in the presence of confining walls differs dramatically from that in the bulk due to hydrodynamic interactions between the particle and the walls~\cite{APP2}.
Another interesting application could be for the classification and design of magneto-optical materials, where the polarization of the reflected light is affected by the local magnetic state due to the Kerr effect, allowing for the visualization of magnetic domain structures with the reported imaging technique~\cite{APP3}. 

\section*{Methods}

\paragraph{Optical microscope setup.}
The prototype microscope~\cite{SRMIC_2020}
uses a high magnification objective lens with high numerical aperture (Nikon CFI Plan Apo Lambda 100$\times$/1.45 Oil). An additional magnification lens in front of the camera makes the total magnification equal to 260$\times$. The critical-type illumination system is customly designed and houses a bright blue ($460 \pm 25$ nm) LED light source (Luminus CBT-120). 
The microscope is configured to operate in  reflection mode, thus providing an improved signal-to-noise ratio. It is equipped with a fast 4 megapixel monochromatic camera (Allied Vision Technologies Bonito CL-400B) working at 100 fps. A liquid crystal polarization rotator device (Meadowlark Optics LPR–200) coupled with a static polarization filter allows analyzing the polarization of the scattered light. The sample can be moved in the horizontal plane by  means of a motorized stage (Micos MS-8), while the vertical movement is achieved by displacing the objective lens, along with the optical system as a whole, with a linear stage (Micos UPM-160). Additionally, the microscope is equipped with pneumatic vibration dumpers (Fabreeka PLM 1).

Microscope components are controlled by a workstation (Dell T7500) equipped with a framegrabber (Matrox Radient eCL), a motion control board (National Instruments PCI-7344) and a GPU board (GeForce GTX 780) for accelerated image processing. The LASSO software framework~\cite{LASSO_link,LASSO_article,LASSO_NewHW} provides modules for real-time microscope control, with fully automated image acquisition and data analysis. 

\paragraph{PSF measurement.}
The PSF is the response of an imaging system to a point source or a point-like object. It is a characteristic of an imaging system and it is often regarded as a measurement of its quality and resolution. In space-invariant imaging systems, such as optical microscopes, the image of a complex-shape object is the convolution of its true shape with the PSF. Moreover, in non-coherent systems the image formation is linear in the intensity, meaning that the microscopic image of a set of objects can also be obtained by the convolution with the PSF. Therefore, the PSF image is an important characteristic of the imaging system allowing to improve the image quality and, under certain conditions, also the resolution. 

A sample of 40 nm silver nanoparticles immersed in a gelatin was used to measure the PSF set: an area of 0.4 mm$^2$ containing  4700 nanoparticles was scanned at the microscope, producing eight sets, one for each polarization angle, of  4700 images each. For the set of images associated to the same polarization angle, the value of all pixels with the same $(i,j)$ index was averaged over the whole set. This procedure generates the set of PSF images shown in figure~\ref{fig:Figure2}. As expected, the PSF images are different, each one being slightly elongated in the direction of the polarization.

\paragraph{NIT sample description and exposure details.}
The NIT emulsion has crystals of 44~nm diameter\cite{NIT_Asada}. After the development, a grain takes the form of a randomly oriented  filament of approximately the same volume. The overall density is equal to 3.44~g/cm$^3$. The granularity corresponds to the average distance between crystals, which is equal to (71$\pm$11)~nm\cite{NIT_Asada}. NIT emulsion film was exposed to carbon ion beam of 100 keV at the ion implantation system (NH-20SR-WMH) of the Nagoya University Nano fabrication Platform. The Carbon ion beam was inclined by about 10$^{\circ}$ with respect to the emulsion film surface.

\paragraph{Electronic microscope, NIT sample preparation and acquisition details.}
SEM images were acquired with a field emission gun SEM (FEG–SEM) FEI/ThermoFisher Nova NanoSem 450. The emulsion film was dipped in hexane to remove the immersion oil, then air dried, mounted on a 25 mm SEM aluminum stub and sputtered with a nanometric conductive layer of Au/Pd using a Denton Vacuum Desk V TSC coating system. The acquisition conditions were optimized to prevent excessive charging of the sample and damage of the gelatin/grains. 
Safe conditions were found by comparing images after several scans of the same area (6 kV accelerating voltage, 7500$\times$ magnification, 10 $\mu$s dwell time, at a resolution of 6144 x 4096). 
In order to maximize resolution and contrast, scans were performed at a low working distance of 3.6 mm, using a DBS directional backscattered electron detector. 
To avoid grain damage, the electron beam voltage was reduced to achieve the minimum penetration depth compatible with high-quality imaging of the grains located below the gelatin surface. The others parameters (lower magnification and higher resolution, dwell time, scan interlacing and averaging) were optimized until the second measurement of the same field of view produced visually identical high-quality image.

\paragraph{Data availability statement.} The datasets generated and analysed during the current study are available from the corresponding author on reasonable request.


\section*{Acknowledgements}
This research was carried out in the frame of the STAR Plus Programme, financially supported 
by UniNA and Compagnia di San Paolo

\section*{Author contributions statement}
A.A. conceived the imaging method and developed the software; T.A. produced the emulsion sample and performed measurements at the optical microscope; T.A. and F.B. performed measurements at SEM; A.A. and V.T. performed the image analysis; G.D.L. guided this research. All authors reviewed the manuscript. 

\section*{Additional information}

\textbf{Competing interests:} The authors declare no competing interests. 


\end{document}